\documentclass[12pt]{article}
\usepackage{graphicx,floatflt}
\date{}
\begin{document}

\def\be{\begin{equation}}
\def\ee{\end{equation}}
\def\bea{\begin{eqnarray}}
\def\eea{\end{eqnarray}}
\newcommand{\bear}{\begin{equation}\begin{array}}
\newcommand{\eear[1]}{\end{array}{#1}\end{equation}}
\newcommand{\gtrsim}{\stackrel{>}{\sim}}
\newcommand{\lessim}{\stackrel{<}{\sim}}
\newcommand{\dst}{\displaystyle}
\newcommand{\bm}{\boldmath}
\newcommand{\fr}[2]{\frac{{\dst #1}}{{\dst #2}}}
\newcommand{\mh}{\mbox{$M_H\,$}}
\newcommand{\epe}{\mbox{$e^+e^-\,$}}
\newcommand{\ggam}{\mbox{$\gamma\gamma\,$}}

\newcommand{\egam}{\mbox{$\gamma e\,$}}
\newcommand{\egeh}{\mbox{$e\gamma\to eH\,$}}
\newcommand{\SM}{${\cal S} {\cal M} \;$}
\newcommand{\MSM}{${\cal M} {\cal S}  {\cal M}\;$}
\newcommand{\DSM}{${2\cal D} {\cal S}  {\cal M}\;$}
\newcommand{\CP}{${\cal C}{\cal P}\;$}
\newcommand{\MSSM}{${\cal M}{\cal S} {\cal S}  {\cal M}\;$}
\newcommand{\fn}[1]{\footnote{ #1}}

\title{{\boldmath\bf Higgs boson.
Complete discovery and the window to a New Physics }\\
\em Proc. Lund Workshop, September, 1998}

\author{Ilya F.~Ginzburg\\
Institute of Mathematics.
  630090. Novosibirsk. Russia.\\ E-mail: ginzburg@math.nsc.ru}

\maketitle

\begin{abstract}
1. We discuss the opportunities of complete Higgs boson
discovery at different colliders.  Main conclusion here is:
Photon collider is the best for this goal.

2. We discuss how to use the Higgs boson (after its discovery) as
the window to a New Physics. Main conclusion here is: Higgs boson
production in gluon and photon fusion together with production in
process $e\gamma\to eH$ provides very good window to a New Physics
with the scale $\Lambda\gg\mh$.
\end{abstract}

Among the main goals for future large colliders is the
discovery of Higgs boson. In the Standard Model (\SM) Higgs
boson is responsible for the observed breaking of the basic
$SU(2)\times\, U(1)$ symmetry; the fundamental particles acquire
masses through the interaction with scalar Higgs field. Until
reliable discovery of Higgs we cannot believe that the \SM is
governed our world.

The construction of \SM admits different variants of Higgs
sector. Before symmetry breaking, Higgs fields can be several
(weak) isotopical doublets, triplets etc. (in the \MSSM --- 2
doublets). In the minimal variant (\MSM) this Higgs field is
single weak isodoublet. In our world, after symmetry breaking,
Higgs sector contains one selfinteracting scalar and single free
parameter of theory is its mass \mh. In the 2--doublet model
(\DSM), after symmetry breaking the physical Higgs sector
contains two neutral scalars, one neutral pseudoscalar (\CP odd)
(in some variants with scalar --- pseudoscalar mixing and \CP
violation) and two charged Higgses. There is large variety in
the possible couplings and sets of physical fields here.
Besides, possible strong interaction in Higgs sector can produce
various heavy "mesons" with $J^P\,=\,0^-,\, 1^+,\, 2^\pm,...$,
which don't excluded by data so far. These opportunities should
be excluded by forthcoming experiments to be sure that the
discovered is indeed the Higgs boson of \MSM or it represents
some other model.

Our discussion is preceded brief description of different
colliders for our aims since some features of a Photon and Muon
colliders are badly known in community. Than we go to the
physical problems noted in the abstract.

Through the paper we use the vacuum expectation value of Higgs
field $v=246$ GeV. We denote by $E$ and ${\cal L}$ the beam energy
and annual luminosity of collider  \cite{PDG}.

\section{Colliders}

The {\bf Hadron colliders} are well known. That are
upgraded Tevatron with $E=1$ TeV, ${\cal L}=2$ fb$^{-1}$
 and LHC  with $E=7$ TeV, ${\cal L}=100$ fb$^{-1}$.

{\bf \boldmath The Muon Collider ($\mu^+\mu^-$)} and its first
stage -- {\bf\boldmath First Muon Collider (FMC)} are under wide
discussion now with the project parameters varied quickly. The
most important for our problems is FMC with c.m.s. energy near
\mh, energy spread $100\div 10$ MeV, annual luminosity 0.1
fb$^{-1}$ and beam polarization about 60\%, which can be rotated
relatively easily \cite{Palmer},~\cite{Skr}. At higher energy the
potential of Muon Collider is close to that of $e^+e^-$ Linear
Collider with the same energy.

{\bf\boldmath Next Linear Colliders in \epe\ mode} with
beam energy $E_e=0.25\to 1$ TeV and annual luminosity
${\cal L} =500$ fb$^{-1}$ are discussed in detail in refs.
\cite{SLACDESY}. They will be complexes having $e^+e^-$ mode and
{\bf\bm Photon Colliders (\egam\ and \ggam)} modes --- {\bf PLC}
with following typical parameters \cite {SLACDESY,GKST}({\em
obtained without special optimization for photon
mode}\footnote{One can obtain relatively easily ${\cal
L}_{\ggam}\approx(0.2\div 0.3) {\cal L}_{ee}$.  Here ${\cal
L}_{ee}$ is the luminosity of the initial \epe\ collider. With
special efforts in the stage of damping rings or earlier the
value of ${\cal L}_{\ggam}$ can be even larger than the basic
\epe luminosity. The considered often opportunity to have higher
${\cal L}_{\ggam}$ with nonmonochromatic photon spectrum seems
unsuitable for the physical program of PLC.}):

\begin{itemize}
\vspace{-0.3cm}
\item {\em Characteristic photon energy $E_\gamma\approx 0.8E_e$.}
 \vspace{-0.3cm}
{\em
\item Annual luminosity ${\cal L}\approx 100$ fb$^{-1}$.
 \vspace{-0.3cm}
\item Mean energy spread
$<\Delta E_\gamma>  \approx  0.07E_\gamma$.
\vspace{-0.3cm}
 \item Mean photon helicity $<\lambda_\gamma>\approx 0.95$
with easily variable sign. One can transform this polarization
into the linear one \cite{kotser}.
\vspace{-0.3cm}
\item There are no reasons against observations at small angles}
(except variable details of design).
\vspace{-0.3cm}
\end{itemize}

\section{Complete discovery of Higgs boson}

In this section we consider Higgs boson in \MSM.  Some cloud of
parameters of more complex models (two doublet, \MSSM, ...) will
be covered with the same efforts.  In this respect one should
solve the series of experimental problems:
\begin{enumerate}
\vspace{-0.3cm}
\item {\em To discover something and to measure its mass.}
\vspace{-0.3cm}
\item {\em To test that spin--parity $J^P$ is $0^+$ and to test \CP.}
\vspace{-0.3cm}
\item {\em To measure couplings with other particles.}
\vspace{-0.3cm}
\item {\em To measure total width $\Gamma_{tot}$.}
\vspace{-0.3cm}
\item {\em To measure Higgs selfcoupling constant.}
\end{enumerate}

The potential of large colliders for the discovery of Higgs
boson was studied in many papers. It is reviewed in
refs.~\cite{LC,SLACDESY},\cite{Gun2}--\cite{Zerw}. The most of
properties of Higgs boson will be obtained by the combination of
results from different colliders.

{\em The results obtained at some single collider are preferable
for the discussed problem since an influence of systematical
inaccuracies here can be reduced much more strong. In this
respect, it is useful to compare the potential of different
colliders for these problems.}

In the table 1 we collect main reactions and decay modes,
useful for Higgs discovery. (The brackets [\ \ ] shows decay
modes that seem observable but they don't studied yet.)

The main mechanism of Higgs boson production at hadron colliders
is {\em gluon fusion $pp\Rightarrow gg\to H$}. But the difficulties
with background suppress often to use this channel for Higgs
boson discovery.  So, for the hadron collider just as for the
\epe collider the {\em associative production} of the Higgs
boson in the states $ZH$, $WH$, $t\bar{t}H$, ...  is considered
mainly.  The missing mass method in the reaction $\epe\to HZ$ is
noted below as MM.  For the \ggam and $\mu^+\mu^-$ colliders we
consider the resonance production of Higgs boson ($\ggam\to H$
or $\mu^+\mu^-\to H$) only and show in the Table only the
discovery channels.
\begin{table}[htb]
\begin{center}
\begin{tabular}{||c||c|c||c|c||c||c||} \hline
Collider$\to$& \multicolumn{2}{|c||}{$\bar{p}p,\;pp$}&
\multicolumn{2}{|c||}{\epe}&\ggam&$\mu^+\mu^-$\\ \hline
$M_H$ {\em GeV}\
$\downarrow$&prod.&decay &prod.&decay&decay&decay\\ \hline
&$WH,\;ZH,$&$\tau\bar{\tau}$,&&&&\\
$95 - 130$&$t\bar{t}H,\; gg\to H$&
$b\bar{b}$,
\ggam&$ZH$&MM, $b\bar{b}$&$b\bar{b}$&$b\bar{b}$\\ \hline
$130 - 155$&$gg\to H$,
$WH$,&[$WW^*$],$\tau\bar{\tau}$,&$ZH$,&MM,
$b\bar{b}$,&\multicolumn{2}{|c||} {$WW^*$, $b\bar{b}$, $ZZ^*$}\\
& $t\bar{t}H$
& $ZZ^*$, \ggam& $H\nu\bar{\nu}$&$[WW^*]$&\multicolumn{2}{|c||} {}\\ \hline
$155 - 180$& $gg\to H$& $WW$, $ZZ^*$&
$ZH$, &MM, $[WW]$, &\multicolumn{2}{|c||}{$WW^*$,$ZZ^*$}\\
&&&$H\nu\bar{\nu}$& $ZZ^*$&\multicolumn{2}{|c||} {}\\ \hline
$> 180$& $gg\to H$& $ZZ$& $ZH,\;H\nu\bar{\nu}$&
$ZZ$, MM&\multicolumn{2}{|c||} {$ZZ$, [$WW$]}\\ \hline
\end{tabular}
\caption{\em Production processes and discovery channels at
different colliders.}
\end{center}
\end{table}

\subsection{The separate problems at the discovery of Higgs boson}

1. {\bf The discovery of something and mass measurement} is the
necessary first step. The \MSM gives no definite predictions for
\mh. The vacuum stability of \MSM limits $M_H$ from below by 105
GeV. The modern LEP data together with most probable estimate
from the high--precision electroweak data gives for \MSM
$90<M_H<310$ GeV~\cite{PDG}. Higher values of \mh are also not
completely excluded. Contrary to that, if it will be found that
$\mh>130$ GeV, the \MSSM is almost excluded.

The interval $M_H\leq 95$ GeV will be covered at LEP2 in nearest
future. The upgraded Tevatron + LEP2 promise to cover mass
interval until $M_H = 130$ GeV.

For $M_H<135$ GeV main decay channel is $H\to b\bar{b}$. It is
expected to find such Higgs at Tevatron or \epe\ linear collider
via the associative production or \ggam\ decay channel for gluon
fusion. The $\tau\bar{\tau}$ channel seems useful for
LHC. The PLC and FMC have also high discovery potential in the
$b\bar{b}$ channel.

If $M_H>2M_Z$, Higgs can be discovered at all colliders via
the sizable decay mode $H\to ZZ$. A background to this decay
mode is rather small.

In the mass range $M_H=140-190$ GeV the decay mode $H\to W^+W^-$
with real and   virtual $W$   ($W^*\to q\bar{q}, e\bar{\nu}, ... $)
is dominant, branching ratios of other decay modes become small,
and their using for the Higgs boson discovery is difficult. The
use of the $H\to W^+W^-$ decay at $e^+e^-$ collider is also
difficult due to a strong nonresonant $WW^*$ background. The
Higgs boson with the mass $M_H=140-190$ GeV can be discovered at
the PLC via the $H\to WW^*$ decay mode \cite{GIvan} with high
efficiency. The potential of $ZZ^*$ channel here should be also
considered.

The FMC provides opportunity to find Higgs boson mass with very
high accuracy that determines by energy spread of collider. The
price for this opportunity is the fact that this collider can
find Higgs boson with $M_H<200$ GeV only {\bf after} discovery
of some signal at another collider.

2. {\bf\bm Testing of spin--parity $J^P$ and $CP$ parity}.
Besides the expected \MSM Higgs boson we can meet axial \CP odd Higgs
partner from \DSM\  or \MSSM\  and some resonances from
possible strong interaction in  Higgs sector with $J^P=0^\pm,\;
1^\pm,\;2^\pm$.  In any case, if   $J^P\neq
0^+$ for the founded particle, it is not the Higgs boson.
If the observed particle is pseudoscalar and it is decoupled
with gauge bosons, one can hope that it is pseudoscalar Higgs
partner from      \DSM or \MSSM.

{\em The \epe and hadron colliders.} When consider Higgs boson
associative production the $ZZ$ final state gives quite
different angular distribution than $ZH$. So, one can be sure
that we cannot mix up H and Z. To exclude spin states like $J^P=
1^+,\,2^\pm$ with similar angular dependencies at least large
additional luminosity integral is needed.  Besides, in $ZH$,
$WH$ channels the angular distributions relative to $Z$, $W$ is
changed dramatically with variation of the parity of produced
scalar.

 For the higher values of \mh correspondent increasing of
necessary energy of \epe collider is necessary and the
observation process $\epe\to HZ$ is changed to W--fusion
$\epe\to\nu\bar{\nu}H$. In this case testing of $J^P$ for the
discovered particle needs   difficult analysis of distributions
of decay products.

{\em The PLC and FMC} allows the best opportunities to test spin
and parity of produced particle via variation of initial
polarization state of system.

Provided the luminosity distribution near the peak is the
Lorencian one, the cross section of the (pseudo)scalar resonance
production, averaged over luminosity distribution, is ($C=\ggam$
or $\mu^+\mu^-$) (sign $"+"$ corresponds scalar, sign $"-"$
corresponds pseudoscalar).

\bea
&<\sigma_C>=\int\sigma(s)\fr{1}{\cal L}\fr{d{\cal L}}{ds}ds \equiv
B_C\cdot \fr{4\pi\Gamma_{H\to A}}{M_H^3}\left[1+\lambda_1\lambda_2\pm
D_\bot\right],&\nonumber\\
&B_C=\fr{1}{\pi}\fr{M_H}{\Gamma_H+\Delta E};&\label{basic}\\
&4\pi\fr{\Gamma_{H\to\mu\mu}}{M_H^3}=\fr{m_\mu^2}{2 v^2M_H^2},\quad
4\pi\fr{\Gamma_{H\to\ggam}}{M_H^3}
=\left(\fr{\alpha}{4\pi}\right)^2\fr{|\Phi|^2}{4 v^2}\quad
(|\Phi| \sim 1);&\nonumber\\
&D_\bot =\ell_1\ell_2 \cos 2\phi\;\mbox{ for \ggam},\quad
D_\bot =4\vec{s}_{1\bot}\vec{s}_{2\bot}\;\mbox{ for
}\mu^+\mu^-\,. &\nonumber
\eea
Here $\Delta E$ is the initial energy spread, $\lambda_i$ are
the helicities of photons or doubled helicities of muons,
$\ell_i$ are degrees of photon linear polarization and $\phi$ is
the angle between them, $\vec{s}_{i\bot}$ are transverse
components of muon spin\footnote{ Note that $\alpha/4\pi\sim
m_\mu/v$. So the difference between averaged cross sections at
PLC and FMC is given mainly by the collider quality factor
$B_C$.}.

Therefore variation of sign of helicity of one photon or muon
switches on or off the production of "Higgs boson". Such
experiment excludes completely states with $J^P=1^\pm,\,2^\pm$.
The variation of linear (transverse) polarizations switches on
or off particle production depending on its parity.  The
necessary luminosity integral here is the same as for the
discovery of the particle. The PLC is better for these problems
due to possible earlier beginning of operations and higher
degree of beam polarization expected.

3. {\bf Coupling of Higgs boson with other particles}.  The
Higgs mechanism of particle mass origin in the \MSM means that
the coupling constant of Higgs boson with the fundamental
particle $f$ is given by the ratio of its mass to the vacuum
expectation value of Higgs field $v$ (for fermions),
$g_f=m_f/v$. If the observed value will be different, then
either some model with many Higgses (e.g. \MSSM) is valid or the
mechanism of mass origin in \SM differs from that considered in
this model.

To measure these couplings (their ratios) one should
resolve Higgs boson signal in both the dominant decay mode and
in another modes. The nonresonant production of these final
states gives very high background. This background is suppressed
in the resonance production of narrow Higgs boson with not too
high mass. Here the position of FMC seems the best.

Good potential (which is not studied in detail yet) has here the
PLC.  For example, about 10 fb$^{-1}$ is necessary to see decay
$H\to ZZ^*$ at $M_H>120$ GeV \cite{GIvan}. So, one can find the
ratio $g_{HZZ}/g_{HWW}$.  If Higgs mass is within interval
125--150 GeV, one can measure additionally $g_{Hb\bar{b}}/
g_{HWW}$, etc. This interval is the best for testing of many
coupling constants of Higgs boson at PLC.

The Table 1 shows the potential of different colliders in
this problem. The measuring of separate coupling constants and
Higgs selfcoupling are very difficult problems at \epe\ or
proton colliders. It needs very high luminosity integral.

4. {\bf The total Higgs width $\Gamma_H$} in the \MSM is
calculated with high accuracy via $M_H$ and the set of
fundamental particle masses. $\Gamma_H< 10$~MeV at
$M_H<140$~GeV, $\Gamma_H<10$ GeV at $M_H>310$~GeV,
$\Gamma_H/M_H>1/4$ at $M_H>700$~GeV.

\begin{floatingfigure}{7cm}
   \centering
 \includegraphics[width=7cm]{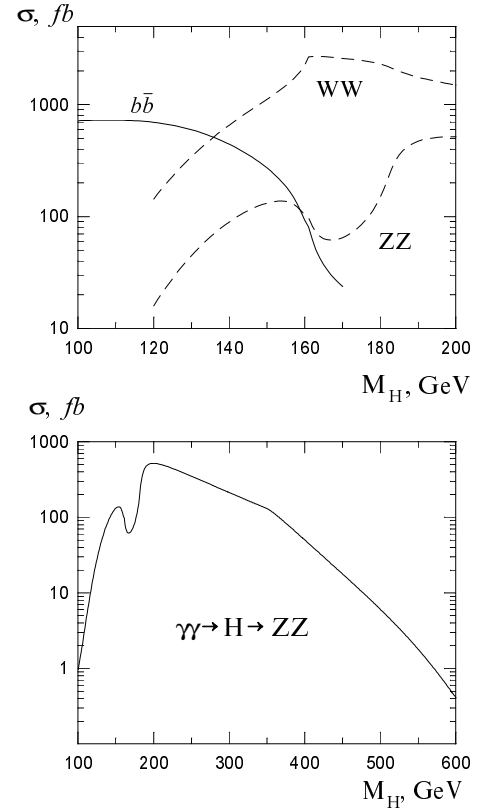}
\caption{\em The cross section of reaction $\ggam\to H$ with some
decay channels. $<\lambda_1>=<\lambda_2>=0.9$;
$20^\circ<\theta<160^\circ$.}
   \label{figggam}
\end{floatingfigure}

One can hope to measure $\Gamma_H$ at $\mh>200$~GeV at all
colliders.  At $M_H<200$ GeV this quantity can be measured
directly at FMC only. In the mass interval $M_H=180-210$ GeV the
cross section of process $\ggam\to WW$ exhibit specific
structure obliged by interference of QED process with Higgs.
With resolution about 5 GeV in the $WW$ effective mass this
structure can be used to both see Higgs boson and to measure its
width.

5. {\bf The Higgs selfinteraction} $\lambda^2 (aH^3+H^4)/8$ is
the basic point of Higgs mechanism.  The \MSM constants are
$\lambda =M_H/v$, $a=2v$. This selfinteraction can be seen in
the processes $\epe\to ZHH$,  $\epe\to\nu\bar{\nu} HH$, $\mu^+\mu^-\to
\nu\bar{\nu} HH$, $\ggam\to WWHH$, $\ggam\to HH$, $\ggam\to HH$, etc. with
cross section about 1 fb or less \cite{Il1,JikHH}.
In these cross sections the diagrams with HHH vertex compensate
strong those without this vertex. So, the result is
sensitive to the possible deviations from above value of this
constant.

{\bf General view.} The expected opportunities to study
different properties of Higgs boson are shown in the Table 2.
When 2 signs are in one square of this table, the upper
corresponds $\mh<200$ GeV, the lower -- 500 GeV$>\mh>200$ GeV.

\begin{table}[htb]
\begin{center}
\begin{tabular}{|p{3cm}|c|c|c|c|}\hline
{\em problem}&Tevatron /LHC&\epe\ collider&Photon
collider&FMC\\ \hline
1 discovery & +&+&+&$\begin{array}{c} -\\
\pm\end{array}$\\ \hline
Mass& +&+& +&The best\\ \hline
Spin--Parity &$
\begin{array}{c}\neq Z,\;\neq 0^-
\\\pm\end{array}$&
$\begin{array}{c}\neq Z,\neq 0^-
\\\pm\end{array} $&\multicolumn{2}{|c|}{ $
J^P\neq 0^+$ excluded} \\ \hline
$\Gamma_{tot}$& $\begin{array}{c}\mbox{---}\\+\end{array}$
&$\begin{array}{c}\mbox{---}\\+\end{array}$ &
$\begin{array}{c}\mbox{Poor}\\+\end{array}$ & The best \\ \hline
Couplings   &$\pm$& $\pm$&$\pm$&
$\begin{array}{c}+\\ \pm\end{array} $\\\hline
Selfinteraction& Poor& $\pm$& $\pm$& $\pm$\\ \hline
\end{tabular}

\caption{\em Potential of Higgs boson complete discovery.}

\end{center}
\end{table}

One can conclude that the PLC and FMC are the best in complete
discovery of Higgs boson with some advantages for PLC. The
analysis of the data in Z peak and estimates in \MSSM show that
the most probable mass interval for Higgs boson is $100\div 200$
GeV. So, as it was proposed earlier, the PLC with the c.m.s.
energy 100--200 GeV open the shortest way to find Higgs boson
and to test its properties completely~\cite{BalGin}.

To estimate necessary luminosity integral $\int{\cal L}$ for the
discovery of Higgs boson at PLC we assume sequence of runs
covered about 10\% of energy interval near the upper bound. Fig.
\ref{figggam} represent the Higgs boson production cross sections for
different essential channels.

The first discovery of Higgs here demands less than 3 fb$^{-1}$
for suitable energy interval\footnote{ If $M_H<145$ GeV, the
discovery of Higgs boson via reaction $\ggam\to b\bar{b}$
demands $\int{\cal L}<3$ fb$^{-1}$ \cite{BBC}. If $135<M_H<195$
GeV, the discovery of Higgs boson via reaction $\ggam\to WW^*$
demands $\int{\cal L}<1$ fb$^{-1}$ \cite{GIvan}. If $M_H>185$
GeV, the discovery of Higgs boson via reaction $\ggam\to ZZ$
demands $\int{\cal L}<1$ fb$^{-1}$.} Therefore, to cover mass
interval 100--200 GeV one should have $\int{\cal L}<15$
fb$^{-1}$.  After discovery of Higgs peak, the same luminosity
integral (no more than 3 fb$^{-1}$) is necessary to test spin
and parity of this particle via the change of polarization (see
eq. (\ref{basic})).

In the mass interval 200--300 GeV main discovery channel is
$\ggam\to ZZ$ with $\int{\cal L}\approx 1$ fb$^{-1}$ for each
energy interval. So $\int{\cal L}<5$~fb$^{-1}$ is necessary to
cover whole discussed region. The additional integral about 2
fb$^{-1}$ is necessary to test spin and parity here.

At $\mh>350$~GeV Higgs signal should be resolved at the {\em
regular} loop induced background $\ggam\to ZZ$. Here the
luminosity integral about $10\div 20$~fb$^{-1}$ is sufficient to
see signal from the Higgs boson (which will be obtained at LHC
before these measurements).

\section{ Higgs window to a New Physics}

Going to higher energies we hope to meet       New Physics with
characteristic scale $\Lambda>1$ TeV. Before discovery new heavy
particles inherent this New Physics, it reveals itself at lower
energies as some {\em anomalies} in the interactions of known
particles.  The goal of studies in this field is to find these
anomalous interactions and discriminate as better as possible
the type of underlying theory via comparison of different
anomalies.

{\bf\bm After discovery of Higgs boson the study of gluon
fusion at hadron collider, processes $\ggam\to H$ and \egeh at
photon colliders will provide the best place for discovery and
discrimination of New Physics effects.} Indeed,

{\em\begin{itemize}
\vspace{-0.3cm}
\item The $Hgg$, $H\ggam$, $H\gamma Z$ vertices are absent in
the \SM  Lagrangian. They are one--loop effects. Therefore, the
relative contribution of anomalies is enhanced here in
comparison with other interactions.
\vspace{-0.3cm}
\item If New Physics adds new heavy particles into the \SM,
their effects are enhanced in the discussed vertices due to
absence of decoupling here.
\vspace{-0.3cm}
\item Gluon or photon fusion are dominant mechanisms of Higgs
boson production at hadron or photon colliders.
\end{itemize}}

The gluon fusion at Hadron Collider and photon fusion at PLC are
studied good in literature. The \egeh process provides
opportunity to study $HZ\gamma$ interaction after the study of
$H\ggam$ interaction in the process $\ggam\to H$.

It is useful to describe briefly why decoupling is absent in
this problem. It is very simple and well known (see e.g.
\cite{Okun}). The Higgs two--gluon decay is the loop effect with
quarks in loop. The contribution of each quark in the amplitude
is proportional to $g_q\alpha_s/max\{M_H,\,m_q\}$. The
Higgs--quark Yukawa coupling constant here $g_q=m_q/v$($m_q$ is
the quark mass).  Therefore, at $m_q<\mh$ this contribution is
small and at $m_q>\mh$ each quark gives the same contribution
independent on its mass. This vertex counts number of quarks
that are heavier than the Higgs boson. The same is valid for
the $H\ggam$ and $HZ\gamma$ interactions, but here the charged
leptons and gauge bosons also contribute in the amplitude, and
contributions of fermions and gauge bosons have opposite sign.
That is why Higgs boson fusion is very sensitive to the new
heavy particles having Higgs mechanism of mass origin. (If the
mass of heavy particle arises not via Higgs mechanism, it
is decoupled with the relatively light Higgs boson.)

Below we discuss some scenarios of New Physics from this point of
view.\\

\subsection{Search of fourth generation \cite{GISch}}

The \SM does not fix the number of generations of quarks and
leptons. There are no ideas, why there is more than one
generation and what Law of Nature limits their number.

The studies at Z peak at LEP closed the window for the 4--th
generation quarks and leptons with light neutrino.  However, the
extra generations with heavy neutrino does not excluded. The
above data shows that if this generation exists, the quarks and
leptons within should be much heavier than 200 GeV and neutrino
is heavier than 45 GeV.

Fortunately, {\em one can answer definitely about existence of
these families before the discovery their very heavy members}
via the study of gluon and photon fusion of the Higgs boson at
proton and photon colliders.

Due to absence of decoupling here, one can consider for the
gluon fusion at Tevatron and LHC only the effects of $t$--quark
and 4--th generation. Then for the Higgs with mass less 200 GeV
the amplitude increases by a factor 3 and the two--gluon width
-- by a factor 9. At higher Higgs masses more detail effects
near $t\bar{t}$ threshold become essential.

\begin{figure}
 \includegraphics[width=0.47\textwidth,height=7cm]{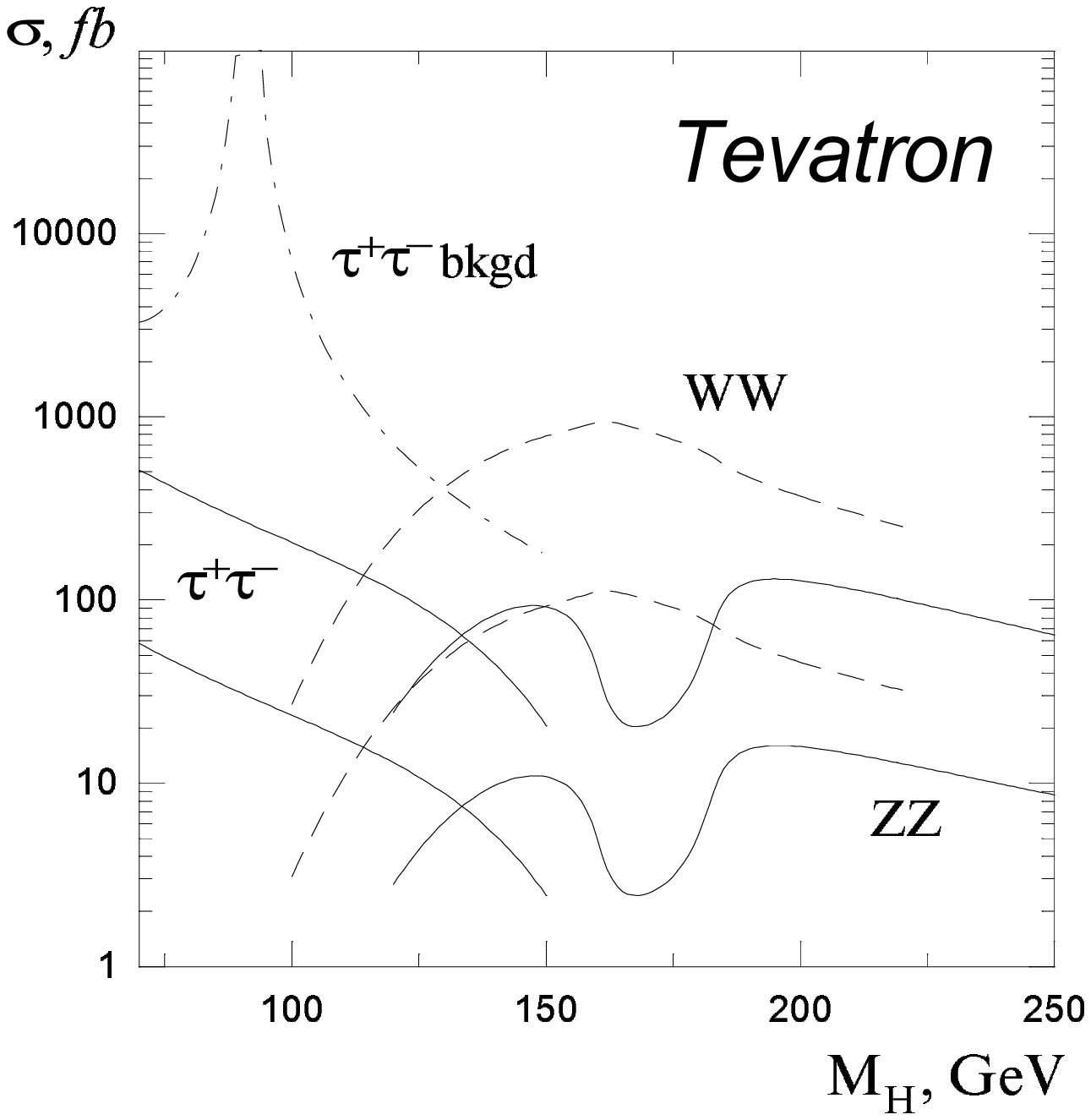}
\hfill
   \includegraphics[width=0.47\textwidth,height=7cm]{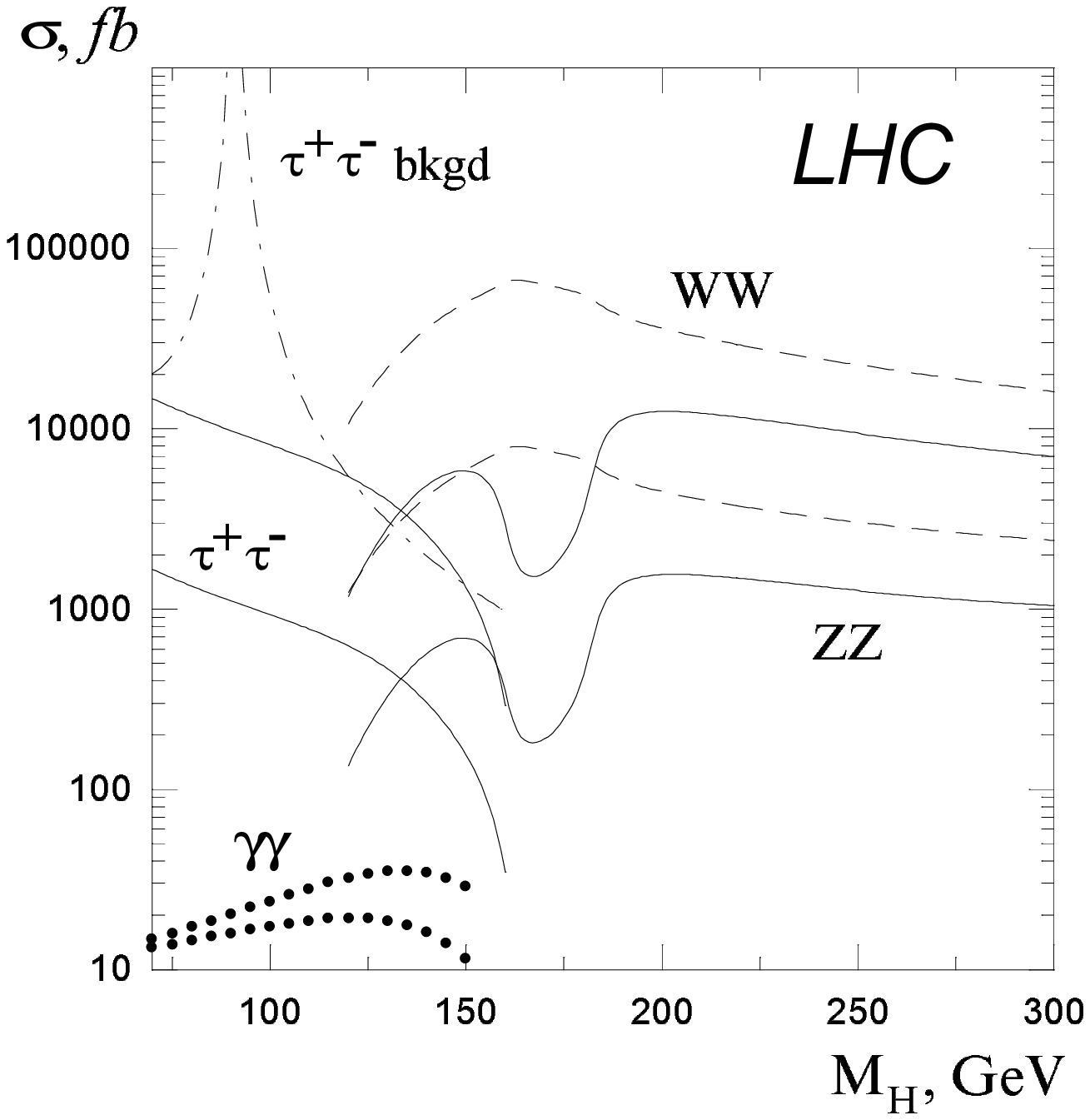}
\caption{\em The cross sections of Higgs boson production at Tevatron
and LHC for different decay channels.  The lower curves --- three
generations, the upper ones --- four
    generations. The $\tau\bar{\tau}$ background is also shown.}
   \label{figglfus2}
\end{figure}

The experimental cross sections is the convolution of the cross
section of the fusion Higgs production with gluon structure
functions. The variations in the production cross sections for
the different decay channels obtained with the structure
functions \cite{strf} shown in the Fig.
\ref{figglfus2}. Here two gluon width and production cross
section in the model with n generations are labeled by
superscript $n$\footnote{  Here only one--loop are presented.
The QCD radiative corrections enhance cross section (but no
background) by a factor about 2. Radiative corrections due
to strong Yukawa interaction of Higgs field with heavy matter
was not considered.}.

\begin{floatingfigure}{0.4\textwidth}
   \includegraphics[width=0.4\textwidth]{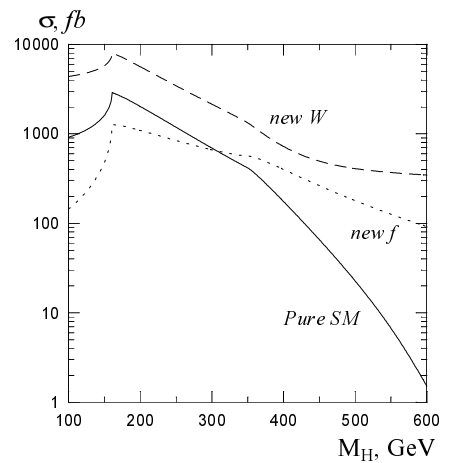}
\caption{\em The effect of new particles within \SM on $\ggam \to
H$ cross section. $<\lambda_1>=<\lambda_2>=0.9$;
$20^\circ<\theta<160^\circ$.}
   \label{figggam3}
\end{floatingfigure}

It is seen that the effect at proton collider is huge.  This
Figure shows the opportunity to find effect of 4--th generation
in different channels at hadron colliders\fn{ The radiative
corrections enhances these cross sections strongly.  The
detector efficiency effects reduce these cross sections.
But their relations are independent from detail of
recording.}. In respect of studies at relatively light Higgs
boson we compare here also the Higgs boson production in
$\tau\bar{\tau}$ channel and the corresponding background
($q\bar{q}\to Z\to\tau\bar{\tau}$). One can see that the signal
from fourth generation should be seen at upgraded Tevatron in
$WW^*$ channel till 190 GeV. The studies of $\tau\bar{\tau}$
production at effective masses below 150 GeV and of $WW^*$
production above 135 GeV could reveal the existence of the
fourth generation even without discovering the Higgs boson in
the associative production.  Studies at LHC cover whole possible
set of the \mh values.

The discussed large effect for gluon fusion
can be imitated in two Higgs doublet
models or in \MSSM at $\tan\beta\approx 20$. To exclude this
explanation, the two photon production of Higgs boson at PLC
should be considered. In this reaction contributions of fermion
and $W$ loops are of opposite sign below corresponding
thresholds. It is the reason why cross section of photon fusion
is decreased strongly at $\mh<300$~GeV as compare with standard
case. In this case 2-doublet model imitate effect at the
different value of $\beta$. Therefore to measure Higgs boson
production at photon collider is necessary to confirm that
effect is connected with new generation. The additional
measurements of \egeh reaction are also useful for this goal.
Figs. 3 and 4 shows effect of new heavy generation in these
processes.

The effect of possible new heavy $W$ -- like boson in the Higgs
boson production at PLC is also shown in Fig. \ref{figggam3}. To
verify this mechanism, additional measurements of \egeh process
are necessary (this new $W$ gives no contribution in the gluon
fusion.)

\subsection{\bm Two doublet \SM -- \DSM}

The simplest extension of \MSM --- \DSM --- contains two doublet Higgs
sector. Let us remind
that its real Higgs sector contains 2 neutral scalars $h$
(light) and $H$ (heavy), one neutral pseudoscalar $A$ and two
charged Higgses $^\pm$ and some set of Yukawa couplings with the
matter fields. The key parameters of model are two mixing
angles $\beta$ and $\alpha$. We assume below the widely
discussed Model II for the Yukawa coupling of Higgs fields with
the matter fields.

The observation of coupling of Higgs boson with other particles
different from that predicted by \MSM would be strong argument
in favor of \DSM. However, it is difficult to expect complete
enough set of observed couplings to conclude about realization
of \DSM definitely.  The discovery of all Higgs particles would
be essential step in this problem.

We discuss the nearest opportunities related to the case of \DSM
in which the argumentation in favor of model different from \MSM
cannot be obtained. That are the cases when either the lightest
Higgs $h$ (pseudoscalar $A$) cannot be seen or observed $h$
imitate Higgs boson of \MSM. In both cases we assume masses of
other Higgses to be much higher than $M_h$.

A. If $\sin^2(\beta-\alpha)<0.1$, the relatively light $h$ or
$A$ ($2\div 50$ GeV) can exist but elusive for modern
experimentation. There is wide enough field of parameters for
this elusive Higgs boson \cite{Krawcz1}. This boson can be seen
either in the processes like $\epe\to t\bar{t}H$, $\ggam\to
t\bar{t}H$ (at large enough $\cot\beta$).

The reliable method to see such Higgs boson is to study it via
loop obliged processes where effects of different quarks
supplement each other. The first idea was to use for this goal
photon fusion \cite{Krawcz1}. However, the necessary low energy
photon collider doesn't considered now as realistic future
machine. The other way is to use process $\egam\to eh$  or
$\egam\to eA$ at \egam collider with c.m.s. energy 100-400 GeV
\cite{GIKr1}. One can expect here effect about 10 fb or larger
\cite{BGIv1}.\\

B. Another opportunity is \DSM, imitating \MSM. It corresponds
$\sin(\beta   -\alpha)=1$. In this case all couplings of $h$ with the
matter are the same as in \MSM. For the first glance, there is
no opportunity to exclude this variant before discovery of
$A,\;H,\;H^\pm$. Fortunately, the coupling with photons is
sensitive to the existence of heavy charged Higgs bosons. So,
the comparative study of gluon fusion and photon fusion together
with \egeh process ($HZ\gamma$ vertex) will resolve such \DSM
from \MSM \cite{GIKr2}.\\

The next problem is -- {\bf\bm how to distinguish \DSM from
\MSSM before discovery (heavy) superpartners}. This problem
become essential after obtaining of some experimental
argumentation against \MSM. Since the variety of possible
parameters of \DSM is large, the solution of this problem can be
obtained through comparison of couplings of Higgs bosons with
gluons and photons (perhaps for all three, $h$, $H$ and $A$)
\cite{GIKr2}.

\subsection{\bm Anomalous interactions in Higgs boson
production at \ggam and \egam collisions}

At energies below scale of New Physics $\Lambda$ this New
Physics manifests itself as some ({\it anomalies}) in the
interactions of known particles. It can be described by
an effective Lagrangian which is written as expansion in
$\Lambda^{-1}$ (For the observable effects it means expansion in
$(E/\Lambda)$):
\be
L_{eff}=L_{SM}+\sum\limits_{k=1}^\infty\Delta L_k\,;\;\; \Delta
L_k =\sum\limits_r \fr{d_{ rk} {\cal O}_{rk}} {\Lambda^k}\,,\;\;\;
(dim\{{\cal O}_{rk}\}= 4+k\,). \ee
Here $L_{SM}$ is the Lagrangian
of the \SM. Below, as usual,
$$
B_{\mu\nu}=\partial_\mu B_\nu -
\partial_\nu B_\mu\,, \;\; W^i_{\mu\nu} =\partial_\mu
W^i_\nu-\partial_\nu W^i_\mu -g\epsilon^{ijk} W^j_\mu
W^k_\nu\,;\;\;
 \tilde{V}_{\mu\nu}
=\fr{1}{2}\epsilon_{\mu\mu\alpha\beta}V_{\alpha\beta}\,.
$$

We have $\Delta L_1=0$. Therefore, we consider the next
largest term $\Delta L_2$. For our problem contributions of
different items ${\cal O}_{r2} $ are joined in the
\be
\Delta L
_v = (2Hv+H^2)\left(\theta_\gamma\fr{F_{\mu \nu}
F^{\mu\nu}}{2\Lambda_{\gamma}^2} +
i\tilde{\theta}_\gamma\fr{F_{\mu \nu } \tilde{F}^{\mu\nu}}
{2\Lambda_{P\gamma}^2}+
 +\theta_Z \fr{Z_{\mu \nu }F^{\mu \nu}}
{\Lambda_Z^2} +i \tilde{\theta}_Z \fr{Z_{\mu \nu }
\tilde{F}^{\mu \nu}}{\Lambda_{PZ}^2}\right)   \;\;\;[\theta_i =\pm 1]\,
.\label{L1}
\ee
The relation of these new scales with those of initial
Lagrangian is obtained easily.

It is useful to write the necessary item in the effective
Lagrangian for Higgs field interaction with electromagnetic
field (including the \SM contribution) in the form:
\be
{\cal L}_{H\ggam}=\fr{\left[ G_\gamma
HF^{\mu\nu}F_{\mu\nu} +i \tilde{G}_\gamma
HF^{\mu\nu}\tilde{F}_{\mu\nu}\right]}{2v}\,,\;\;
{\cal L}_{HZ\gamma}=\fr{\left[G_Z H F^{\mu\nu}Z_{\mu\nu}
+i \tilde{G}_Z HF^{\mu\nu}\tilde{Z}_{\mu\nu}\right]}{v}\,.
\ee
with coupling "constants"
\be
G_i=G_i^{SM} +\theta_i\fr{v^2}{\Lambda_i^2}\,,\quad
\tilde{G}_i=\tilde{G}_i^{SM}+\tilde{\theta}_i\fr{v^2}
{\Lambda_{Pi}^2}\,.
\ee
The \SM values of these couplings are well known (see e.g.
\cite{Gun2,Zerw}).

In the \MSM we have $\tilde{G}_i^{SM}=0$. In the extended \SM
(e.g. in \DSM) scalar and pseudoscalar Higgs field can be mixed,
in this case additionally $Im \tilde{G}_i^{SM}\neq 0$ at
$M_H>2M_t$. \\

\begin{figure}
   \includegraphics[width=0.47\textwidth,height=8cm]{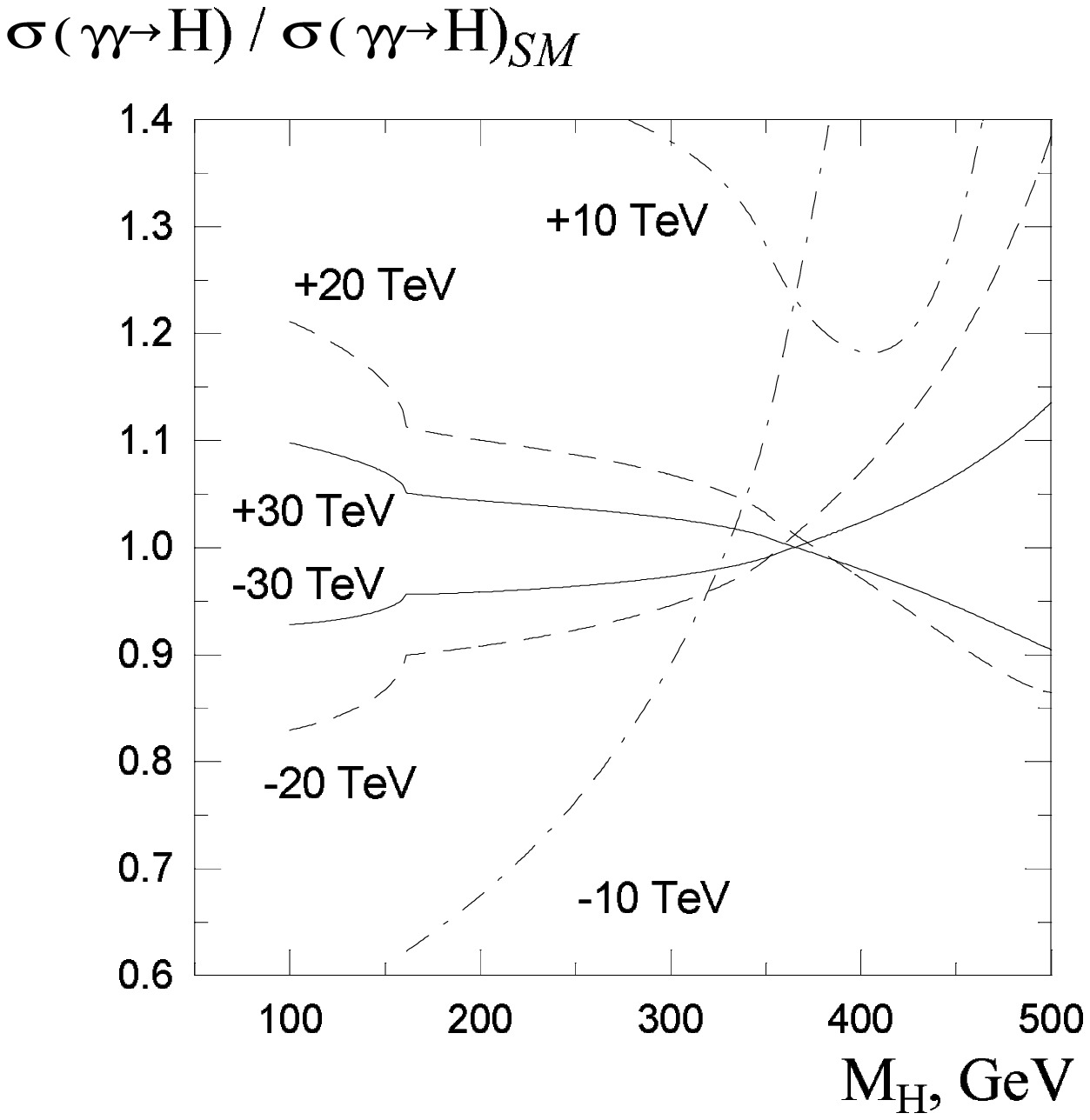}
\hfill
   \includegraphics[width=0.47\textwidth,height=8cm]{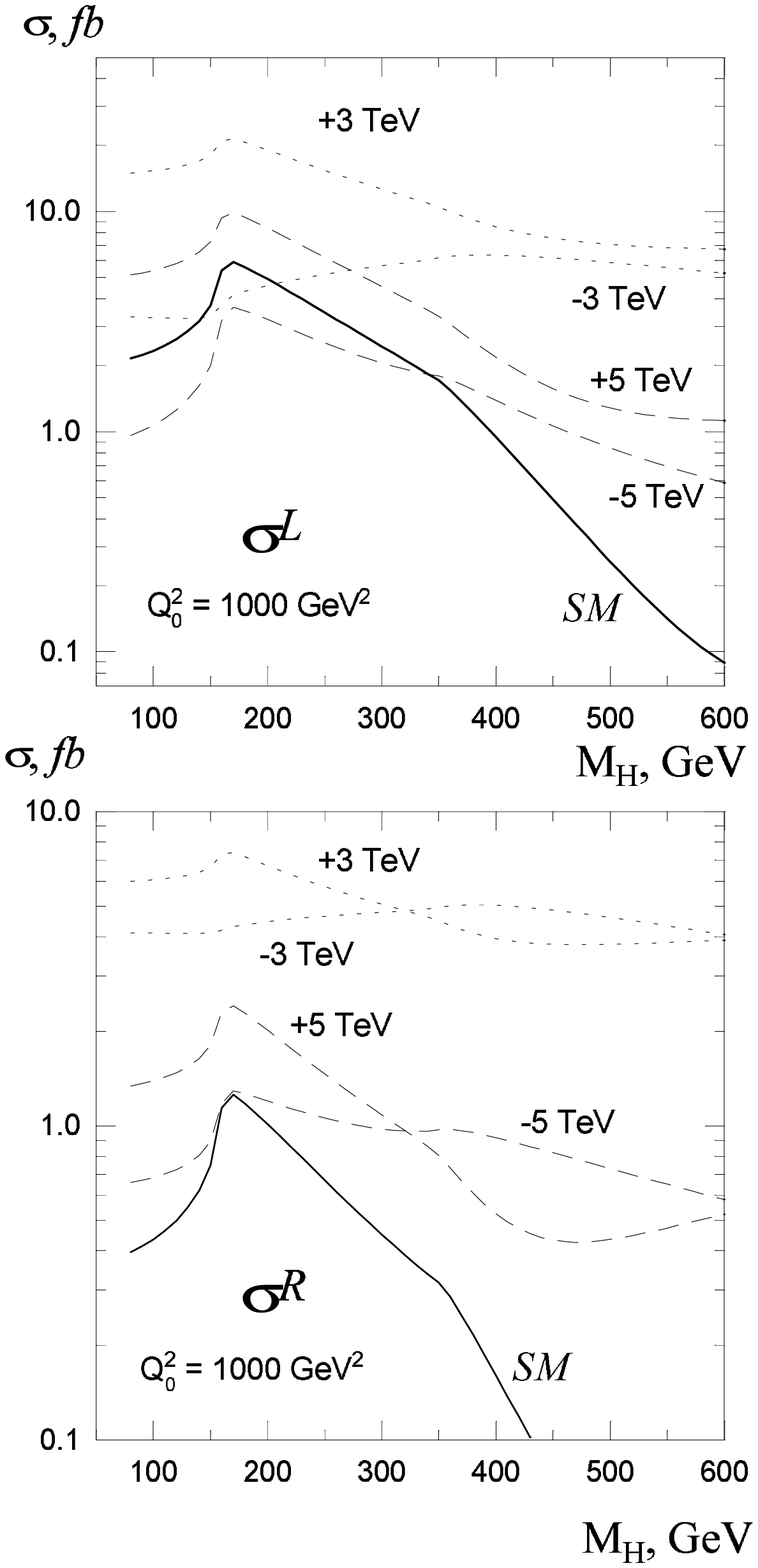}
\caption{\em The modification of the $\ggam\to H$ cross section caused by
$H\ggam$ anomaly and
cross sections  $\sigma_L$ and $\sigma_R$ of \egeh process
caused by $Z\gamma H$ anomaly. The numbers
denote $\theta_i\Lambda_i$. }
   \label{figegehlam}
\end{figure}

{\bf\bm The process $\ggam\to H$ provides the best place for the
study of $H\ggam$ anomalies} \cite{GinH,BGIv1,BGIv3}. The cross
section of process averaged over the photon spectrum is (cf.
(\ref{basic} and notations there):
\bear{c}
 <\sigma>= <\sigma>_{np}^{SM}\fr{T(\lambda ,\phi)}
{|G_\gamma^{SM}|^2}\,,\\
T(\lambda ,\phi)= \left(|G_\gamma|^2
+|\tilde{G}_\gamma|^2\right)(1+\lambda_1\lambda_2)
+ \left(|G_\gamma|^2 -|\tilde{G}_\gamma|^2\right)\ell_{T1}\ell_{T2}
\cos 2\phi \\
+2Re(G^*_\gamma\tilde{G}_\gamma) (\lambda_1 +\lambda_2)
+2Im(G^*_\gamma\tilde{G}_\gamma)\ell_{T1}\ell_{T2}\sin 2\phi
\,.\\
\eear{\label{cpoddggam}}
The effect of \CP even anomalies (neglecting \CP odd) is shown
in Fig. \ref{figegehlam}. It is seen that the two--photon
production cross section is sensitive to the anomalies with the
scale $\Lambda= 20-25$ TeV for the most probable Higgs boson
mass.

The earlier estimates of possible sensitivity to the \CP odd
anomalies without detail study of interference are incorrect
since effect $|\tilde{G}_\gamma|^2\propto 1/\Lambda^4$ is
of the same order of value as the contribution of 8-th order \CP
even operator in $|\tilde{G}_\gamma|^2$ item.

It is seen that the interference \CP odd item with the basic \CP
even one should be clearly seen via variation of cross section
with variation of signs of both photon helicities simultaneously
(both left $\Rightarrow$ both right) and via the study of
production cross section dependence on the angle between
directions of linear polarizations $\phi$.\\

{\bf\bm The process \egeh provides the best place for the
study of $HZ\gamma$ anomalies} after the study of $H\ggam$
anomaly in the process $\ggam\to H$ \cite{GinH,BGIv1,BGIv3}.

Let us start from description of \SM case \cite{BGIv1,Il2}. We
consider the process amplitude only for real ingoing photon. It
is a sum of three items. The first one, ${\cal A}_\gamma$, is
the $\gamma$ exchange contribution (photon exchange between
scattered electron and triangle loop describing $\gamma^*\gamma
\to H$ subprocess).  This item is evidently gauge--invariant
since longitudinal item in photon propagator gives in the
electron vertex $q^\mu u(p')\gamma^\mu u(p)\to u(p')
(\hat{p}-\hat{p}') u(p)=0$. The second item is the $Z$ --
exchange contribution ${\cal A}_Z$ ($Z$--boson exchange between
scattered electron and triangle loop describing $Z^*\gamma\to H$
subprocess). This item is {\em approximately} gauge--invariant
with accuracy $\sim m_e/M_Z$. The residual item (we denote it as
box) is, consequently, gauge--invariant with this very accuracy.
Calculations show that its contribution is small in comparison
with each pole contribution.

The effects of photon and $Z$ exchange are dominant in the total
cross section.  The numerical analysis shows that the effects of
$Z$ exchange become close to that of photon exchange in the
cross section integrated over the region of the transverse
momenta of scattered electrons $p_\bot\ge p_{\bot 0}\approx 30$
GeV. Besides, these contributions almost compensate each other
in the cross section for the right hand polarized electrons and
they interfere constructively. The corresponding cross section is
about 10 fb.

These figures show the opportunity to see effect of $HZ\gamma$
interaction in the \egeh process.

{\bf\bm Effect of $HZ\gamma$ anomalies} in \egeh process is
studied in \cite{BGIv1,BGIv3}. The opportunity to see effects of
\CP even anomalies (neglecting \CP odd) is seen from Fig. \ref{figegehlam}. It
is seen that one can expect to see anomalies with the scale
$\Lambda\approx 4-5$ TeV.

The effect of \CP odd anomalies can be studied via the
polarization dependence of cross section which reproduces main
features of cross section (\ref{cpoddggam}). To understand the
result one should take into account that the linear polarization
of virtual photon or $Z$ is directed in its scattering plane and
its circular polarization is about $(M_h^2/s)2\Lambda_e$ where
$\lambda_e$ is helicity of electron \cite{GinSer80}. Therefore,
the effect of \CP odd anomalies can be seen via the study
dependence of cross section on the sign of photon helicity and
on the angle between electron scattering plane angle relative to
the direction of photon linear polarization.

The possible limits for new \CP odd and $HZ\gamma$ effects will
be enhanced if some anomalous $H\ggam$ effects exist. \\

{\em I am thankful to I. Ivanov and M.Krawczyk for discussions.
This work was supported by grant RFBR 96-02-19079.}

\end{document}